\documentstyle[aps,pre,epsf]{revtex}

\begin{document}
\draft  % makes pacs numbers print

\tighten
\firstfigfalse
\twocolumn[\hsize\textwidth\columnwidth\hsize\csname@twocolumnfalse\endcsname

\title{Probability distribution of the order parameter
       for the 3D Ising model universality class:
       a high precision Monte Carlo study}
\author{M.~M.~Tsypin}
\address{Department of Theoretical Physics, Lebedev Physical Institute,
         117924 Moscow, Russia}
\author{H.~W.~J. Bl\"ote}
\address{Faculty of Applied Physics, Delft University of Technology,
         P.O. Box 5046, 2600 GA Delft, The Netherlands; and Lorentz
         Institute, Leiden University, P.O. Box 9506, 2300 RA Leiden,
         The Netherlands}

%\date{September 23, 1999}
\maketitle

\begin{abstract}
We study the probability distribution $P(M)$ of the order parameter
(average magnetization) $M$, for the finite-size systems at the
critical point. The systems under consideration are the 3-dimensional
Ising model on a simple cubic lattice, and its 3-state generalization
known to have remarkably small corrections to scaling. Both models
are studied in a cubic box with periodic boundary conditions.
The model with reduced corrections to scaling makes it possible
to determine $P(M)$ with unprecedented precision. We also obtain
a simple, but remarkably accurate approximate formula
describing the universal shape of $P(M)$.
\end{abstract}


\bigskip
]

This work is devoted to the study of the following problem.
Consider a finite system belonging to the universality class of the 
three-dimensional (3D) Ising model, exactly at its critical
point. Let the system have a non-conserved order parameter, cubic
symmetry and periodic boundary conditions.
For such a finite-size system the order parameter $M$ (for the
Ising model, the sum of all spins, divided by the total number
of spins in the system) will be a fluctuating quantity,
characterized by the probability distribution $P(M)$ 
\cite{Binder81,Bruce81}.
In the scaling limit (system size going to infinity) this function
is universal (up to rescaling of $M$) and can be thus considered
a very interesting and informative characteristic of the given
universality class. (One should bear in mind that $P(M)$ depends
on the geometry of the box, and on the boundary conditions;
in this study we always consider a cubic box with periodic
boundary conditions). For example, $P(M)$ contains the information
about {\em all} momenta $\langle M^k \rangle$ of $M$, including
the universal ratios such as the Binder cumulant 
$U = 1 - (1/3) \langle M^4 \rangle / \langle M^2 \rangle^2 $,
which has been a subject of many Monte Carlo studies 
\cite{Binder81,BPTR85,Bhanot86,LaiMon,FeLa91,KimPa93,BLH,Ballest99,Ballest98}.
Moreover, the precise knowledge of $P(M)$ proved to be important
for locating and characterizing the critical point in Monte Carlo 
studies of various systems, including the liquid-gas critical point
\cite{Wilding}, the critical point in the unified theory of
weak and electromagnetic interactions \cite{ourEWPT} and in quantum
chromodynamics \cite{ForcrandQCD}.

The first Monte Carlo computation of $P(M)$ for the 3D Ising model
in a cubic box with periodic boundary conditions has been performed in
Ref.~\cite{Binder81}, where its double-peak shape was established.
A more accurate determination of $P(M)$ has been done in
Ref.~\cite{HilferW}, also by Monte Carlo. Results reported for the 3D
case in Ref.~\cite{Stauffer98} appear to be incorrect.

Despite considerable progress in computation of
$P(M)$ by analytical methods \cite{BrezinZJ,RGJ,Eisen,Rudnick98,ChenDohm}, 
numerical simulation remains the main source of information about its 
properties.

Our aim was to compute $P(M)$ on a qualitatively new level of accuracy,
in comparison to what has been done before \cite{HilferW}, and to put
the result into form convenient for further use. We would like to
emphasize the following two features of our computation that made this
possible.

1. The computation of Ref.~\cite{HilferW} used the 3D Ising model on a simple
cubic lattice of the size $20^3$ and $30^3$. As we will see, the shape 
of $P(M)$ obtained with these relatively small lattice sizes still
differs noticeably from its scaling limit, due to nonnegligible
corrections to scaling. To get over this difficulty, we employed,
besides the 3D Ising model, the more sophisticated model in the
same universality class, which was shown to have remarkably small
corrections to scaling \cite{BLH}. This made it possible to 
determine the scaling limit of $P(M)$ with an accuracy far exceeding 
what would be achievable when one is restricted to the standard 3D Ising
model.

2. The existing results for $P(M)$ were presented in the
form of Monte Carlo-generated histograms \cite{Binder81,HilferW}.
We present a simple 3-parameter formula which is suitable for   
quantitative applications. Its accuracy is about $2 \cdot 10^{-3}$ of
the maximum value of $P(M)$.

We have performed Monte Carlo simulations of two models.
The first one is the standard 3D Ising model on the simple cubic 
lattice, defined by the partition function
\begin{equation} \label{Zisi}
  Z = \sum_{ \{s_i \} } \exp \Bigl\{
  \beta \sum_{\langle ij \rangle} s_i s_j  \Bigr\},
  \qquad s_i = \pm 1.
\end{equation}
Here $\langle ij \rangle$ denotes the pairs of nearest neighbours,
and the sum is over the $2^N$ possible configurations, where $N$ is
the total number of spins. We simulate this model at the critical 
point, which we take to be at $\beta_c = 0.221654$, using the 
Swendsen-Wang cluster algorithm \cite{SwWa}, and lattice sizes ranging
from $12^3$ to $58^3$ (with periodic boundary conditions).

The second model (with dramatically reduced corrections to scaling, as was
shown in Ref.~\cite{BLH}) is the spin-1 Blume-Capel model \cite{Blume,Capel}.
Here the spins can take 3 discrete values: $-1$, 0, and $+1$. 
The model is defined by the partition function
\begin{equation} \label{Zspin1}
  Z = \sum_{ \{s_i \} } \exp \Bigl\{
  \beta \sum_{\langle ij \rangle} s_i s_j - D\sum_m s_m^2  \Bigr\},
  \quad s_i = -1, 0, +1 .
\end{equation}
The sum thus includes $3^N$ possible configurations. The parameter $D$
is fixed to the special value $D=\ln 2$ (as explained in Ref.~\cite{BLH}),
and we perform the simulations at the critical point, which is taken 
to be $\beta_c = 0.393422$ \cite{BLH}, using  lattice sizes from $12^3$
to $58^3$. The simulations used  a hybrid method,
which alternates one Metropolis sweep with 5 or 10 Wolff \cite{Wolff}
steps, depending on the system size, as described in Ref.~\cite{BLH}.

The probability distribution $P(M)$ is obtained as follows.
For each configuration generated by the Monte Carlo algorithm,
we determine the order parameter $M = {1\over N} \sum_{i=1}^N s_i$,
and increment the population of the corresponding bin of the 
histogram by 1.

\begin{figure}

\centerline{
\epsfxsize=9cm \epsffile{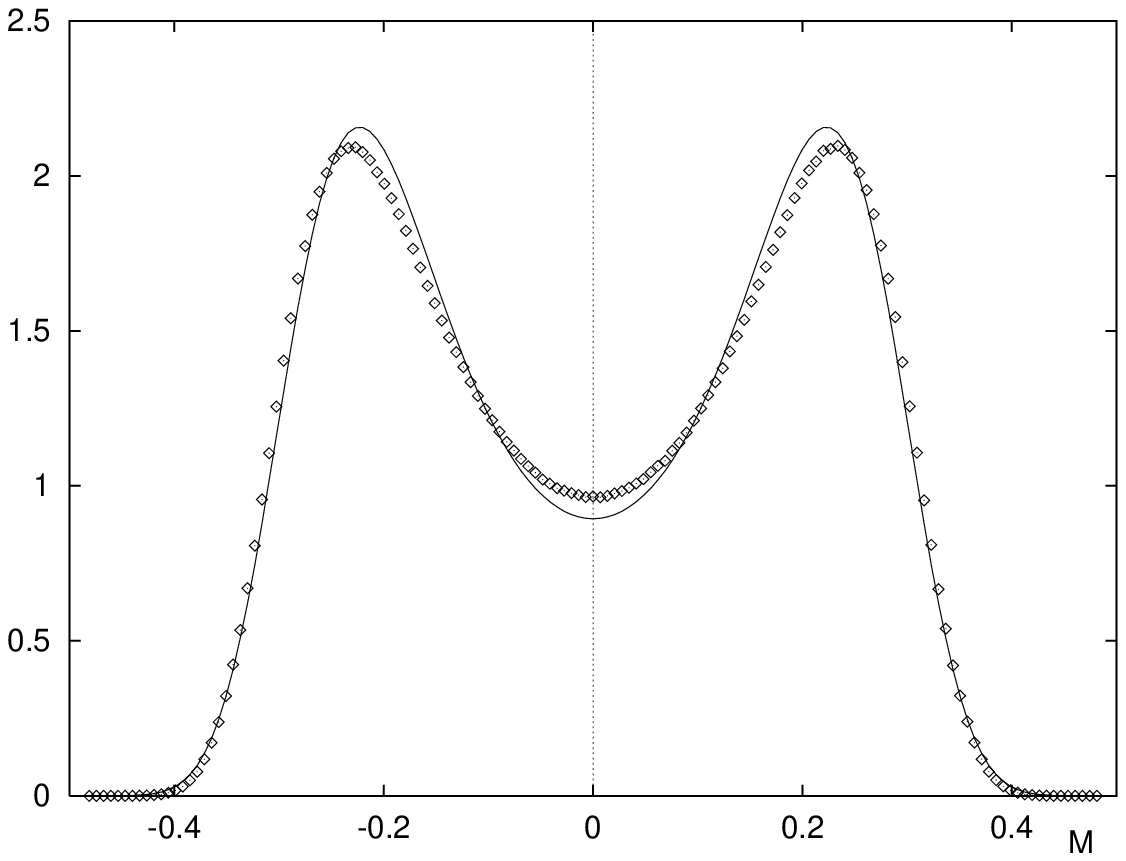}
\epsfxsize=9cm \epsffile{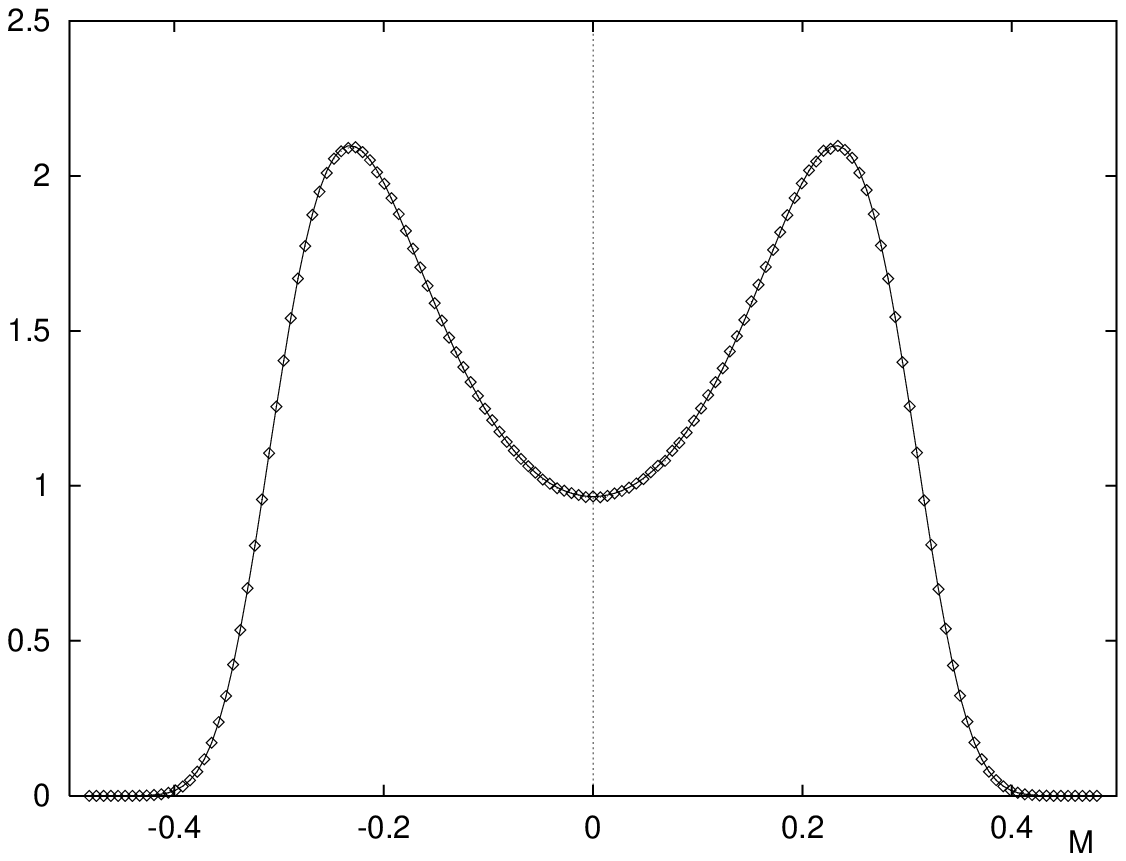}}

\vspace*{0.5cm}

\centerline{
\epsfxsize=9cm \epsffile{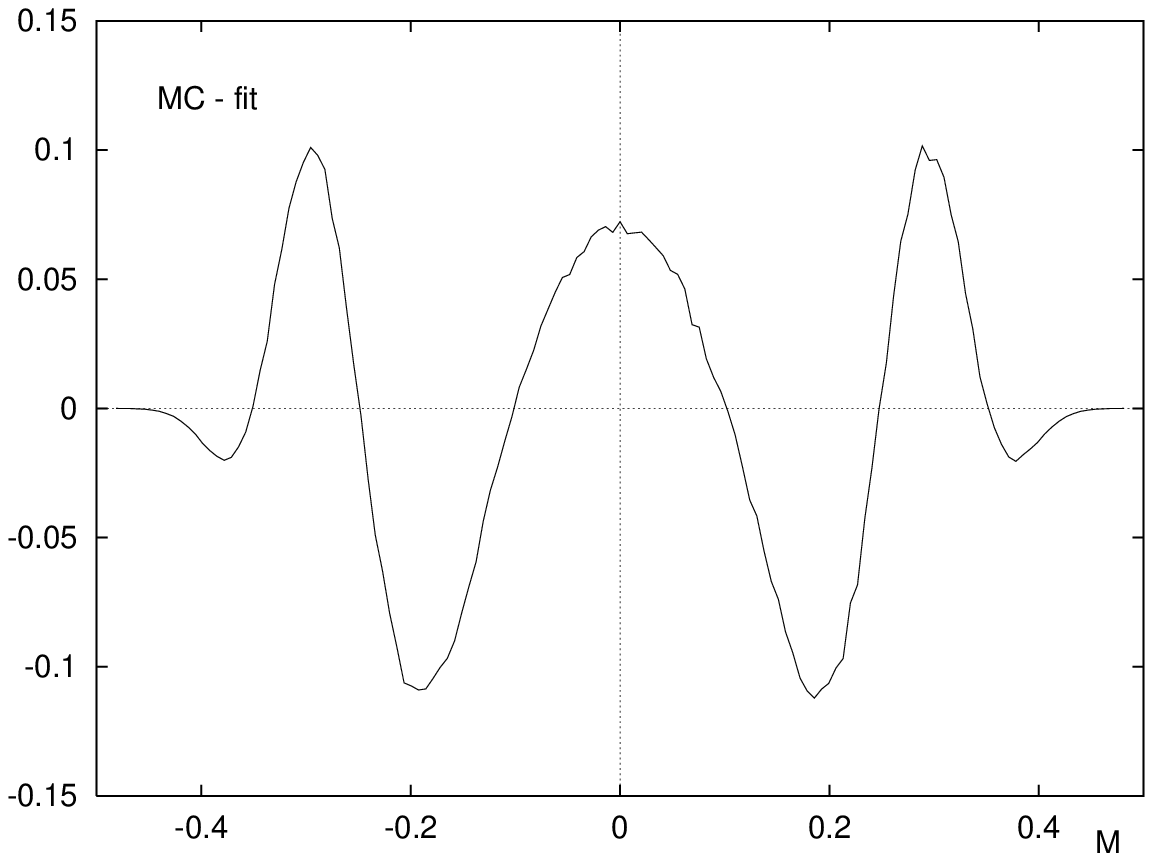}
\epsfxsize=9cm \epsffile{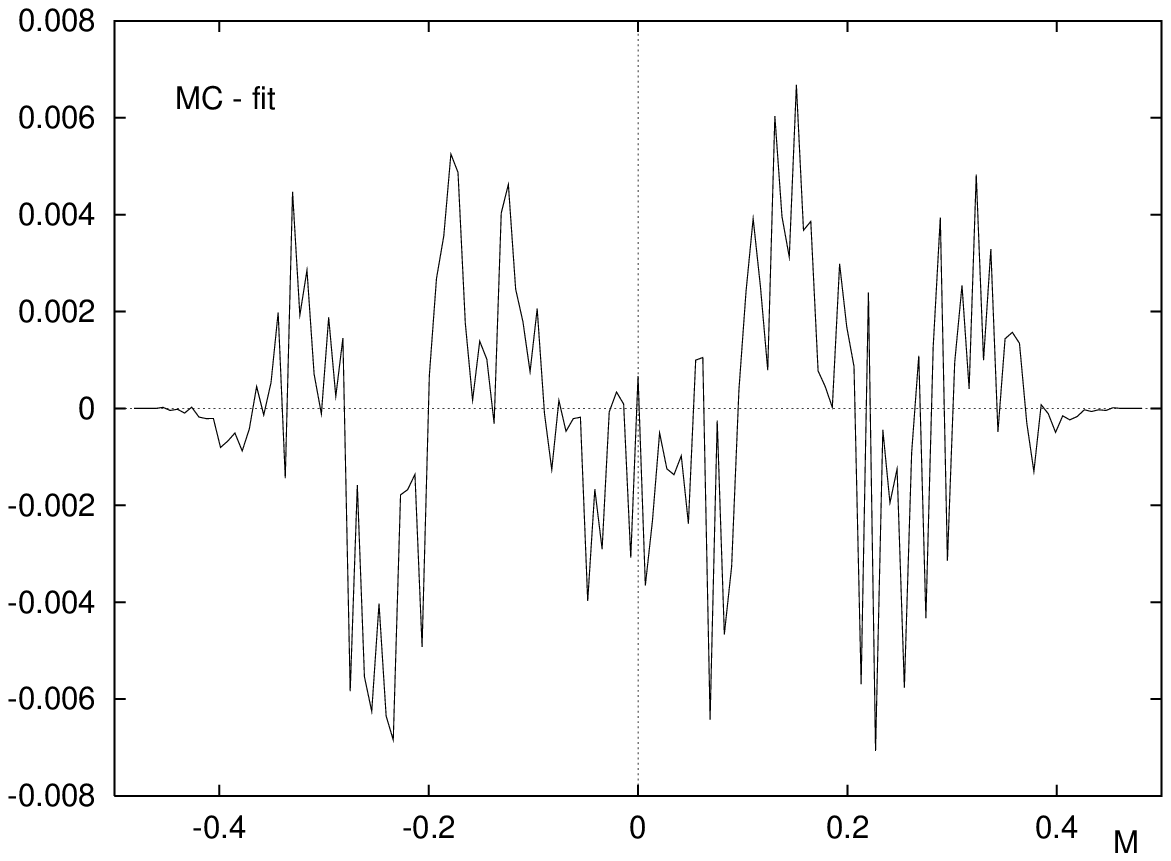}}

\vspace*{0.5cm}

\caption{
Probability distribution $P(M)$ of the spin-1 model with reduced
corrections to scaling defined by Eq.~(\protect\ref{Zspin1}), and
its description by approximations given in Eqs.~(\protect\ref{phi4})
(left) and (\protect\ref{phi6}) (right).
{\em Top}: $P(M)$ obtained by Monte Carlo simulation at the
critical point (diamonds): $20^3$ lattice, $\beta = 0.393422$,
$36 \cdot 10^6$ configurations, 1 Metropolis sweep + 5 Wolff 
steps per configuration. The solid line is the best fit with
Eq.~(\protect\ref{phi4}) (top left) and with 
Eq.~(\protect\ref{phi6}) (top right).
{\em Bottom}: the difference between the Monte Carlo data and 
the fit, corresponding to the plot above it.
}
\label{mc_fits}
\end{figure}

We have found that the following ansatz gives a surprisingly good 
approximation to $P(M)$:
\begin{equation}
 P(M) \propto \exp \Bigl\{  
 - \Bigl( {M^2 \over M_0^2} - 1 \Bigr)^2 
   \Bigl( a {M^2 \over M_0^2} + c \Bigr) 
 \Bigr\} . 
\label{phi6}
\end{equation}
At the same time, the simpler ansatz
\begin{equation}
 P(M) \propto \exp \Bigl\{ 
 - c \Bigl( {M^2 \over M_0^2} - 1 \Bigr)^2 
 \Bigr\}  
\label{phi4}
\end{equation}
is clearly insufficient. This is illustrated in Fig.~\ref{mc_fits},
using our highest-statistics data set for the $20^3$ lattice.
One observes that the accuracy of approximation (\ref{phi6})
is approximately 20 times higher than that of Eq.~(\ref{phi4}),
and the residual discrepancy of Eq.~(\ref{phi6}) is comparable to the
statistical noise, even with the high statistics used.

The ansatz (\ref{phi6}) was motivated by the observation
that $M^6$ plays an important role in the effective potential of the 
models in the 3D Ising universality class, while higher powers
of the order parameter can usually be neglected \cite{TetWett,Tsy1,Tsy2}.
That is, the effective potential can in many cases well be 
approximated by a polynomial consisting of $M^2$, $M^4$ and
$M^6$ terms. This is exactly what appears in the exponent in 
Eq.~(\ref{phi6}). 

The approximate nature of the ansatz (\ref{phi6}) manifests itself 
by its failure to correctly reproduce the large-$M$ behaviour of the
tails of $P(M)$, which is governed by the critical index $\delta$,
\begin{equation}
 P(M) \propto M^{(\delta-1)/2} \exp \{ -const \cdot M^{\delta+1} \}
\end{equation}
(see Ref.~\cite{Bruce95}; for the discussion of the preexponential factor
in a more general setting, see Ref.~\cite{Tsy1}). However, due to the fact
that for the 3D Ising universality class the exponent 
$\delta+1 \approx 5.8$ is close to 6, this does not prevent
the ansatz (\ref{phi6}) from accurately describing the main part
of $P(M)$ (excluding extremely-far-tail region).

The polynomial in the exponent of Eq.~(\ref{phi6}) has three parameters.
Instead of simply parametrizing it by the coefficients in front of
$M^2$, $M^4$ and $M^6$, we have chosen the parametrization so as to
separate the scale-invariant parameters ($a$ and $c$) and the 
scale-dependent parameter $M_0$ (which parametrizes the position of
the peak of the order parameter). The values of $a$ and $c$ in the
scaling limit are universal and determine the ``universal shape''
of $P(M)$.

\begin{figure}

\centerline{ \epsfxsize=12cm \epsffile{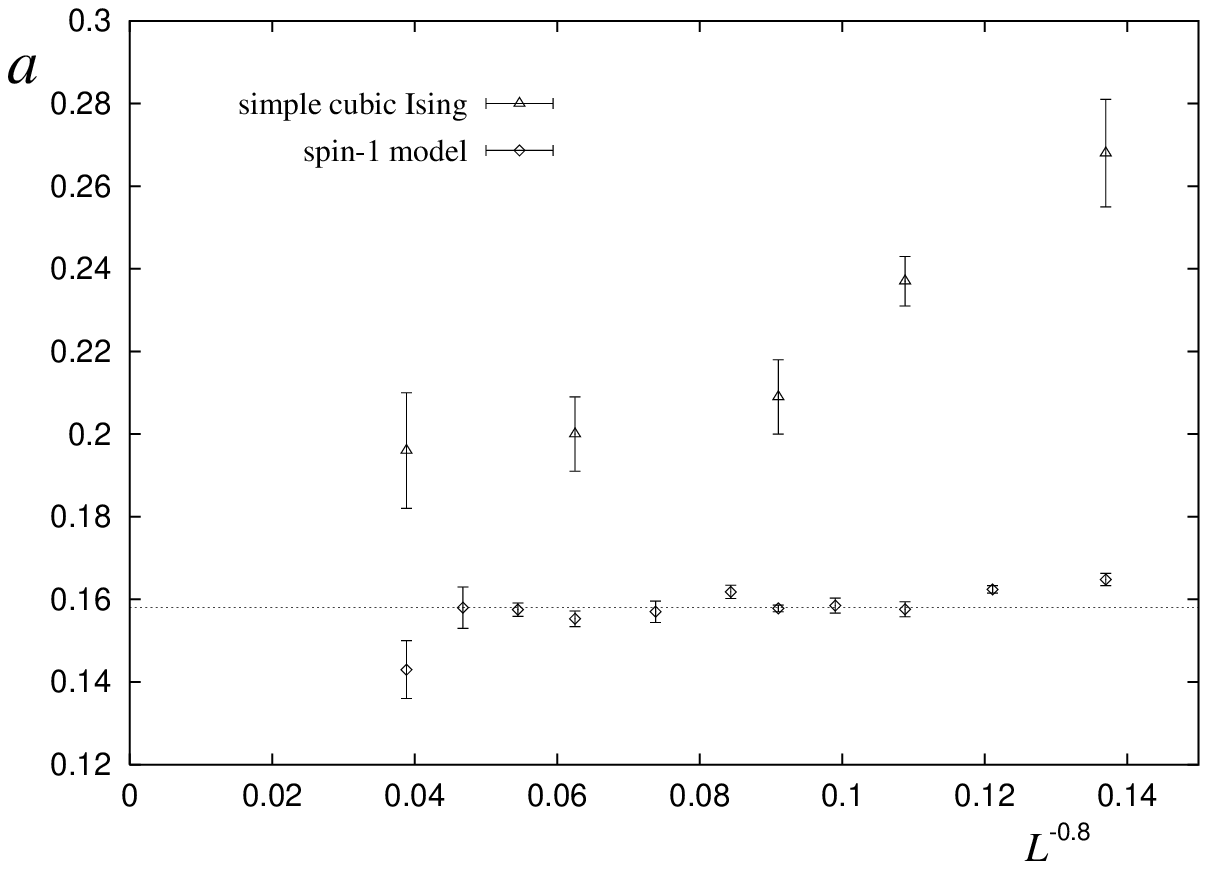} }

\vspace*{0.5cm}

\centerline{ \epsfxsize=12cm \epsffile{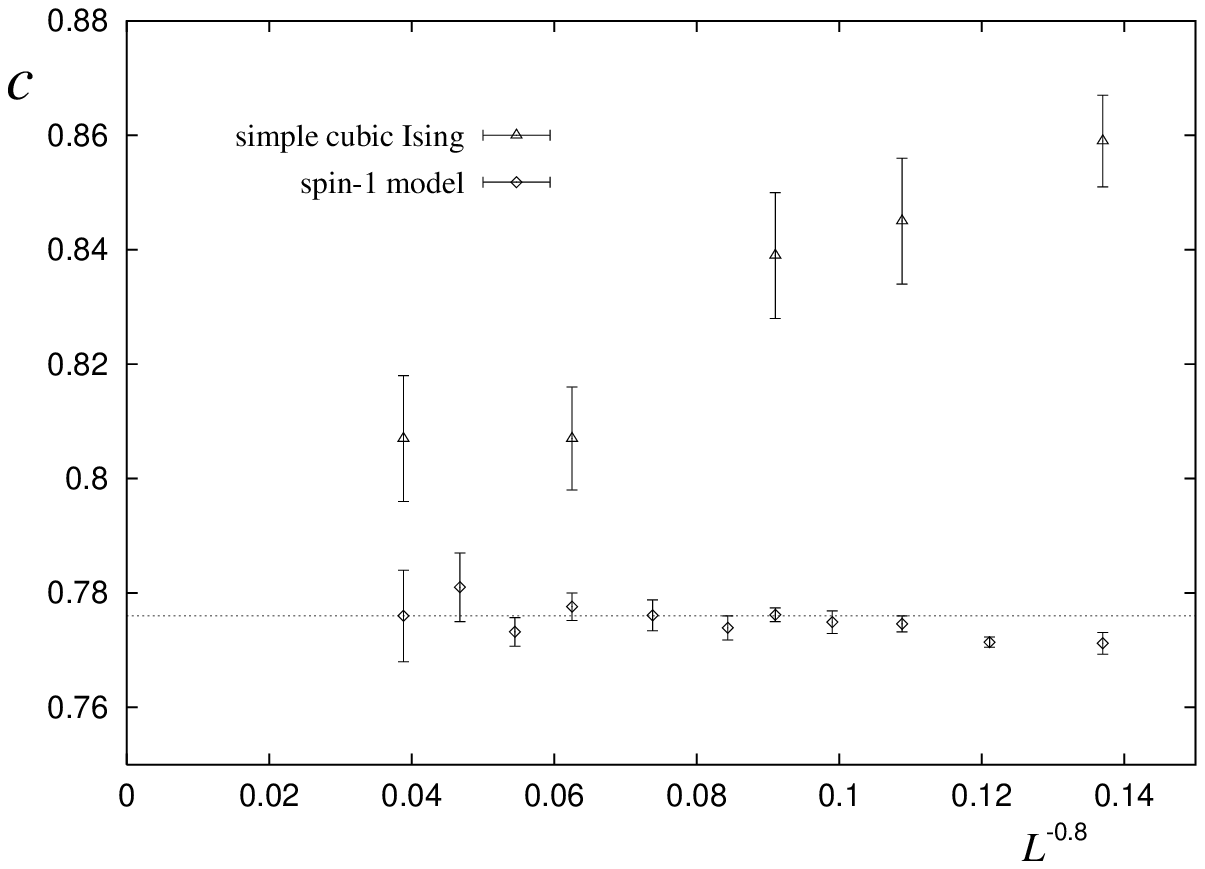} }

\vspace*{0.5cm}

\caption{
Dependence of the scale-invariant parameters $a$ (upper plot)
and $c$ (lower plot) of the probability distribution $P(M)$,
approximated by Eq.~(\protect\ref{phi6}), on the lattice size $L$.
The data for the spin-1 model (diamonds) and for the simple cubic 
Ising model (triangles) are taken from Tables \protect\ref{table1}
and \protect\ref{table2}, respectively. The power of the lattice
size in the horizontal axis is chosen to linearize the leading 
corrections to scaling, which behave as $L^{-\omega}$,
where various estimates give $\omega = 0.80 \pm 0.04$
(see, e.~g., Refs.~\protect\cite{ZJbook} and \protect\cite{GZJ}).
}
\label{ac_vs_L}
\end{figure}

The results of our Monte Carlo simulations are collected in Tables
\ref{table1} and \ref{table2} and shown in Figures \ref{mc_fits},
\ref{ac_vs_L}. For the spin-1 model, no deviations from scaling
are observed on lattices $16^3$ and larger, while the simple cubic
Ising model demonstrates pronounced corrections to scaling, which
are, even on our largest lattices, much higher than both statistical
errors of our spin-1 simulations and the accuracy of approximation 
(\ref{phi6}). Corrections to scaling make it difficult to
extract accurate scaling limit values of $a$ and $c$ from the simple
cubic Ising model data, even if statistical errors are reduced
by a higher-statistics simulation, due to necessity to extrapolate
to $L \to \infty$.

There is no such problem with the spin-1 model, and we obtain the
universal parameters of Eq.~(\ref{phi6}),
\begin{equation}
 a = 0.158(2), \quad  c = 0.776(2).
\label{ac}
\end{equation}
Here the errors take into account both statistical uncertainties
and the systematic deviations inherent in the approximation (\ref{phi6}).
The latter are estimated from the lower right plot in Fig.~\ref{mc_fits}.

{}From Eqs.~(\ref{phi6}) and (\ref{ac}) one can easily obtain any required
property
of $P(M)$. For example, one immediately learns that the ratio of the peak
value of $P(M)$ to its value at $M=0$ is
\begin{equation}
 e^c = 2.173(4).
\end{equation}
Summarizing, we have computed, with a higher accuracy than previously
available, the scaling limit form of the probability distribution $P(M)$
of the order parameter $M$ of systems with 3D Ising universality,
in a cubic box with periodic boundary conditions. A convenient
description of $P(M)$ is given by Eqs.~(\ref{phi6}) and (\ref{ac}),
which deviates from the actual $P(M)$ by no more than $2 \cdot 10^{-3}$
times its maximum value (Fig.~\ref{mc_fits}, right).

\acknowledgements

We thank INTAS (grant CT93-0023) and DRSTP (Dutch Research School for
Theoretical Physics) for enabling one of us (M.T.) to visit Delft
University.

\onecolumn

%%%%%%%%%%%%%%%%%%%%%%%%%%%%%%%%%%%%%%%%%%%%%%%%%%%%%%%%%%%%%%%%%


\newpage

\begin{table}
\caption{
The parameters $a$, $c$ and $M_0$ of the probability distribution
$P(M)$, approximated by Eq.~(\protect\ref{phi6}), obtained by the
Monte Carlo simulation of the spin-1 model defined by
Eq.~(\protect\ref{Zspin1})
at the critical point ($\beta = 0.393422$). M5W means that a new
configuration is produced by one Metropolis sweep followed by 5 Wolff
steps (see Ref.~\protect\cite{BLH} for details). The last three columns 
are: the scale-invariant (but nonuniversal) quantity $M_0L^{d-y_h}$, 
where $y_h = 2.4815(15)$ \protect\cite{BLH,Ballest99},
$\chi^2$, characterizing the quality of fitting the Monte 
Carlo-generated histogram for $P(M)$ by the ansatz (\protect\ref{phi6}),
and the number of bins in this histogram.
}
\label{table1}

\medskip

\begin{tabular}{ccccccccc}
%\hline
Lattice & Method & Configs. & $a$ & $c$ &
 $M_0$ & $M_0L^{0.5185}$ & $\chi^2$ & $N_{\rm bins}$ \\
\hline
$12^3$ & M5W & $10^7$ & 0.1648(15) & 0.7712(19) &
 0.30243(16) & 1.0969(6) & 208 & 116  \\
$14^3$ & M5W & $3.6\cdot 10^7$ & 0.1624(9) & 0.7714(9) &
 0.27915(9)  & 1.0967(4) & 267 & 129  \\
$16^3$ & M5W & $10^7$ & 0.1576(18) & 0.7746(14) &
 0.26035(11) & 1.0962(5) & 175 & 121  \\
$18^3$ & M5W & $10^7$ & 0.1585(18) & 0.7749(20) &
 0.24500(14) & 1.0965(6) & 156 & 125  \\
$20^3$ & M5W & $9\cdot 10^6$ & 0.1568(17) & 0.7782(25) &
 0.23194(15) & 1.0964(7) & 128 & 127  \\
$20^3$ & M5W & $3.6\cdot 10^7$ & 0.1578(8) & 0.7762(12) &
 0.23194(7)  & 1.0964(3) & 218 & 129  \\
$22^3$ & M5W & $10^7$ & 0.1618(16) & 0.7739(21) &
 0.22108(11) & 1.0980(6) & 181 & 125  \\
$26^3$ & M5W & $10^7$ & 0.1570(26) & 0.7761(27) &
 0.20243(14) & 1.0963(8) & 164 & 125  \\
$32^3$ & M5W & $10^7$ & 0.1553(19) & 0.7776(24) &
 0.18180(11) & 1.0965(7) & 166 & 127  \\
$38^3$ & M10W & $10^7$ & 0.1575(16) & 0.7732(25) &
 0.16633(10) & 1.0967(7) & 151 & 128  \\
$46^3$ & M10W & $2\cdot 10^6$ & 0.158(5) & 0.781(6) &
 0.1506(2) & 1.0964(14) & 136 & 123  \\
$58^3$ & M10W & $7.2\cdot 10^5$ & 0.143(7) & 0.776(8) &
 0.1331(3) & 1.0927(25) & 135 & 119  \\
%\hline
\end{tabular}
\end{table}

\begin{table}
\caption{
Analogous to Table~\protect\ref{table1}, but for the Monte Carlo
simulation of the simple cubic Ising model (\protect\ref{Zisi})
at the critical point ($\beta = 0.221654$). SW stands for the
Swendsen-Wang cluster algorithm \protect\cite{SwWa}.
}
\label{table2}

\medskip

\begin{tabular}{ccccccccc}
%\hline
Lattice & Method & Configs. & $a$ & $c$ & 
 $M_0$ & $M_0L^{0.5185}$ & $\chi^2$ & $N_{\rm bins}$ \\
\hline
$12^3$ & SW & $7.2\cdot 10^5$ & 0.268(13) & 0.859(8) & 
 0.3892(11) & 1.412(4) & 129 &105 \\
$16^3$ & SW & $7.2\cdot 10^5$ & 0.237(6) & 0.845(11) &
 0.3360(6)  & 1.415(3) & 81.5 &75 \\
$20^3$ & SW & $7.2\cdot 10^5$ & 0.209(9) & 0.839(11) &
 0.2984(9)  & 1.411(4) & 142 &119 \\
$32^3$ & SW & $7.2\cdot 10^5$ & 0.200(9) & 0.807(9) &
 0.2344(5)  & 1.414(3) & 138 &117 \\
$58^3$ & SW & $7.2\cdot 10^5$ & 0.196(14) & 0.807(11) & 
 0.1733(7)  & 1.423(6) & 121 &119 \\
%\hline
\end{tabular}
\end{table}

\end{document}